# Magnetoelectric effect in hydrogen harvesting: magnetic field as a trigger of catalytic reactions

Donghoon Kim[1], Ipek Efe[2], Harun Torlakcik[1], Anastasia Terzopoulou[1], Andrea Veciana Picazo[1], Erdem Siringil[1], Fajer Mushtaq[1], Carlos Franco[1], Josep Puigmartí-Luis[4,5], Bradley Nelson[1], Nicola A. Spaldin[2], Chiara Gattinoni*[2,3], Xiangzhong Chen*[1], Salvador Pané*[1]

[1] Multi-Scale Robotics Lab, Institute of Robotics and Intelligence Systems, ETH Zürich, Tannenstrasse 3, CH-8092 Zürich, Switzerland. E-mail: chenxian@ethz.ch; vidalp@ethz.ch

[2] Materials Theory Lab, Department of Materials, ETH Zürich, Wolfgang-Pauli-Strasse 27, 8093 Zürich, Switzerland

[3] Department of Chemical and Energy Engineering, London South Bank University, 103 Borough Rd, London SE1 0AA, UK. E-mail: gattinoc@lsbu.ac.uk

[4] Department of Physical Chemistry, University of Barcelona, Martí i Franquès, 1, 08028, Barcelona, Spain

[5] ICREA, Pg. Lluís Companys 23, Barcelona 08010, Spain

**Abstract**

Magnetic fields have been regarded as an additional stimulus for electro- and photocatalytic reactions, but not as a direct trigger for catalytic processes. Multiferroic/magnetoelectric materials, whose electrical polarization and surface charges can be magnetically altered, are especially suitable for triggering and control of catalytic reactions solely with magnetic fields. Here, we demonstrate that magnetic fields can be employed as an independent input energy source for hydrogen harvesting by means of the magnetoelectric effect. Composite multiferroic $CoFe_2O_4$-$BiFeO_3$ core-shell nanoparticles act as catalysts for the hydrogen evolution reaction (HER) that is triggered when an alternating magnetic field is applied to an aqueous dispersion of the magnetoelectric nanocatalysts. Based on density functional calculations, we propose that the hydrogen evolution is driven by changes in the ferroelectric polarization direction of $BiFeO_3$ caused by the magnetoelectric coupling. We believe our findings will open new avenues towards magnetically induced renewable energy harvesting.

**Introduction**

Magnetic fields have been extensively investigated as catalytic reaction boosters to enhance photocatalytic and electrocatalytic performances for clean energy production [1, 2]. Adding magnetic components to current catalysts allows magnetic fields to play a significant role in carrier and mass transportation in catalytic systems and can enhance the efficiency of photo- and electrocatalytic processes [3-5]. For example, magnetic fields can be used to engineer the spin polarization of magnetic catalysts so that the catalysts and chemical adsorbents have coherent spin states, improving electron transfer efficiency between catalysts and chemical adsorbents. This results in faster kinetics and significantly increased catalytic activities in the hydrogen evolution reactions (HER) [6-8] and the oxygen evolution reactions (OER) [9, 10]. However, whether the magnetic field can be employed as the sole trigger of hydrogen energy harvesting is still unanswered despite the advantages of deep penetration depth, low noise and damage, and flexibility in control parameters (i.e. magnitude and frequency) that magnetic fields offer.

Multiferroic and magnetoelectric nanocomposites provide opportunities for exploiting magnetic fields as a direct trigger for hydrogen production [11-14]. While magnetic fields can influence the motion of electrons in magnetic materials, they cannot generate internal electric fields nor charges that are necessary for catalytic reactions. In contrast, magnetoelectric coupling occurs in multiferroic magnetoelectric composite materials when a magnetic field is applied. In typical strain-mediated magnetoelectric composites, the magnetic component responds to magnetic fields and transfers the magnetostrictive strain to the ferroelectric/piezoelectric component by interfacial interaction [15, 16]. The resulting change in electric polarization generates surface charges, which can ultimately induce catalytic reactions if the magnetoeletric material is interfaced with an electrolyte. Here, we demonstrate magnetically induced HER using multiferroic core-shell nanoparticles for the first time. Hydrogen evolution was observed on applying alternating magnetic fields to magnetoelectric nanoparticles dispersed in aqueous solutions. Our first-principles density functional calculations suggest that magnetic field induced changes in the direction of the ferroelectric polarization causes the generation of charge carriers at the surface, which in turn promotes the catalytic reactions.

**Results and discussion**

Here, magnetoelectric $CoFe_2O_4$-$BiFeO_3$ (CFO-BFO) core-shell nanoparticles have been employed as a heterogenous magnetoelectric-HER catalyst. CFO was chosen for the magnetostrictive core because it is known to have very high magnetostriction coefficient (max. 600 ppm) [17, 18]. CFO core particles were synthesized using co-precipitation and hydrothermal methods and coated with a BFO shell using a sol-gel method (see Experimental Methods in Supplementary Information for details). The crystalline structure of the as-synthesized CFO-BFO core-shell nanoparticles was analyzed with X-ray diffraction (Figure 1a). The Rietveld refinement confirms the presence of crystalline CFO and BFO, having cubic Fd-3m and hexagonal R3c space groups, respectively, without any secondary phase (the optimized parameters and the reliability parameters of the Rietveld refinement are provided in Table S1). The high angular annular dark field scanning transmission electron microscopy (HAADF-STEM) and corresponding EDX mappings clearly confirm the core-shell nature of nanoparticles with a CFO core, containing Co, Fe and O, and a BFO shell with Bi, Fe, and O elements. The estimated core size and the shell thickness of the synthesized CFO-BFO core-shell nanoparticles are 35 ± 8 nm and 4 ± 1.5 nm, respectively. As-synthesized CFO-BFO nanoparticles possess good piezoelectric and magnetoelectric properties, which are confirmed by the piezoresponse force microscope (PFM). A representative AFM topography, PFM amplitude, and phase images are provided in Figure S1. The clear phase contrast in Figure S1 indicates different polarization directions existing in the core-shell nanoparticles. Under the application of 50 mT DC in-plane magnetic field, local piezoelectric hysteresis loops show noticeable magnetoelectric coupling in CFO-BFO nanoparticles, as evidenced by the large shifts of positive and negative coercive voltages from 2.61 V to 1.51 V and from -3.88V to -3.04 V, respectively. In addition, the piezoelectric response also increased under the magnetic field (Figure 2). We propose the quantity $\alpha_E = \Delta E/\Delta H$ as a measure of the magnetoelectric coupling in the nanoparticles, where $\Delta H$ is the increment of the external magnetic field and $\Delta E$ is the increment of the induced electric field as previously reported [19]. In this case, 50 mT (500 Oe) of external magnetic field induced an electric field of (1.10 V – 0.84 V)/2/8nm = 16.25 MV m$^{-1}$. Therefore, the calculated magnetoelectric coefficient of the CFO-BFO core-shell nanoparticle is $\alpha_E = 32.5 \times 10^4$ mV cm$^{-1}$ Oe$^{-1}$, which is comparable to previously reported values [13].

Magnetically driven HER via the core-shell nanoparticles were then measured using an online gas chromatography setup under 22.3 mT, 1.19 kHz AC magnetic field (details of the measurement setup can be found in Figure S2 and S3). When particles were dispersed in deionized (DI) water (10 mg/10 mL), 4.01 µmol/g of hydrogen was produced after 6 hours (Figure 3a). To reveal the field-dependent performance, experiments were conducted for 8 hours in total and the magnetic field was turned off for 2 hours during which no hydrogen evolution was observed. In addition, we tested control samples under the same magnetic field (DI water, CFO nanoparticles + DI water) and no hydrogen evolution was observed. These results show that the hydrogen evolution is only triggered by the application of a magnetic field to the CFO-BFO core-shell nanoparticles; we therefore attribute the HER observation to the magnetoelectric effect. The low yield of the evolved hydrogen may be accounted for by rapid electron-hole recombination [20, 21]. To test this possibility, we then added methanol to the DI water as methanol has been used as a reactive species scavenger to enhance the hydrogen production yield [22, 23]. It has been proved by using methanol/D$_2$O isotope water that methanol molecules do not react directly to produce hydrogen but rather consume holes generated in the catalysts [24]. As expected, with the addition of methanol (methanol : DI water = 1 : 9), 28.7 µmol/g of hydrogen was produced after 6 hours (Figure 3a). The effect of the magnetic field intensity and frequency on hydrogen evolution was further investigated in a methanol/DI water solution. In the given range, hydrogen evolution increased when higher intensities and frequencies were applied (Figure 3b and 3c).

We now examine the possible mechanisms for the magnetoelectrically induced HER. When a magnetic field is applied, magnetostrictive CFO responds to the magnetic field and gives strain to the BFO shell. This in turn induces changes in the electronic structure and polarization of ferroelectric BFO, which likely promote the catalytic reactions on the BFO surface [25]. To investigate the electronic structure of strained BFO surfaces and their role in catalytic reactions, we performed first-principles calculations based on density functional theory (DFT). Without loss of generality, we studied a [001] oriented slab of BFO, consisting of stacks of positively charged $Bi^{3+}O^{2-}$ and negatively charged $Fe^{3+}O^{2-}_2$ layers (Figure 4a). As a result of the charged layers, BFO has unstable charged (001) surfaces [26] and the ferroelectric polarization in such a slab aligns itself so that it provides negative (positive) charges on the positive BiO (negative $FeO_2$) surface, providing full compensation of the surface charges (Figure 4a) [25, 27]. This electrostatically stable combination of surface terminations and polarization directions results in no electric field across the slab, with the energies of the band edges independent of their layer position in the slab (Figure 4a). Slabs containing BFO (111) surfaces show the same behavior (Figure S4).

We began by imposing strain by varying the in-plane lattice parameters in our calculations and found that the direct effect of strain on the electronic properties of BFO slabs is small as in the case of bulk BFO [28], with negligible change to the magnitude of the polarization and the electronic structure (Figure 4a, top), for strains of up to $\pm 3$ % (The maximum strain imposed by the CCFO is about 0.06%). We further calculated the adsorption and dissociation energy on water on strained BFO and observed very small changes with respect to the unstrained case (Table S2). Therefore, another mechanism has to be responsible for the increased catalytic activity under magnetic field of the CFO-BFO nanoparticles. It was recently reported that the reversal of the ferroelectric polarization direction has a strong effect on electronic properties [25]. When the polarization direction is reversed, the positive (negative) end of the polarization terminates at the positive BiO (negative $FeO_2$) surface, leading to a severe band bending (Figure 4a, bottom). As a result, the valence band maximum of the positive BiO surface shifts above the fermi level, and the conduction band minimum of the negative $FeO_2$ surface shifts below the fermi level, 'providing' charges at both surfaces. To verify that electrons are indeed generated on the CFO-BFO nanoparticle surfaces under magnetic field, we dispersed CFO-BFO nanoparticles in $AgNO_3$ solution and applied an AC magnetic field. After the field application, Ag nanoparticles were deposited around the CFO-BFO nanoparticles, which clearly indicates the generation of electrons at the BFO surfaces (Figure S5). Based on Mott-Schottky measurement, it can be inferred that the conduction band minima of CFO-BFO nanoparticles is located slightly above the hydrogen potential (Figure S6). Therefore, we can conclude that when the direction of polarization reverses and band bending occurs, the electrons generated on the surface of BFO can participate in HER (Figure 4b).

While the change of polarization is a promising explanation of how the catalytic process occurs under magnetic field, it still needs to be established how such changes of polarization can be achieved in the experimental system through the strain generated by the magnetoelectric coupling. We propose three possible mechanisms, shown schematically in Figure 4c: (i) If strain gradients are generated in the BFO shell by the magnetostriction of CFO, the flexoelectric effect could cause local switching of the polarization [29, 30]. (ii) In a polydomain BFO shell, the different orientations of the polarization domains will be next to each other. For example, as shown in the middle of Figure 4c, neighboring patches of BFO (001) with BiO and $FeO_2$ surface termination are likely to occur. By means of strain, the domain walls separating these patches could move and create local polarization reversal [31-33]. (iii) It has been shown that in-plane strain can create a preference for the ferroelectric polarization direction: perpendicular to the surface for compressive strain and parallel to the surface for tensile strain [34]. Rotation of the ferroelectric polarization to the direction parallel to the surface would lead to polar surfaces, since the charge coming from the ionic layers is left uncompensated.

## Conclusion

In summary, we demonstrated the use of a magnetic field as a trigger for renewable energy harvesting by exploiting the magnetoelectric effect via CFO-BFO core-shell nanoparticles that can clearly act as HER catalyst. We propose a mechanism that when a magnetic field is applied, the magnetostrictive CFO core responds to the field and transfers the strain to the BFO shell, causing changes in the BFO ferroelectric polarization direction. The reversal of the polarization caused by the magnetoelectric coupling results in charge generation at the particle surface, which would be the driving force for the magnetoelectrically induced catalytic reactions. We believe our findings will provide opportunities for efficient hydrogen energy production.

# Figures

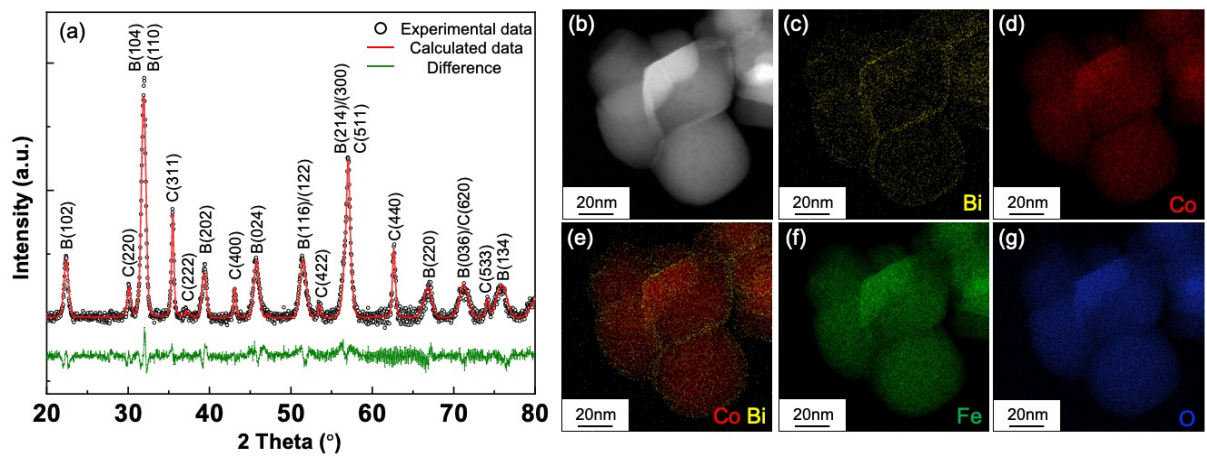

**Figure 1.** (a) Theta-2theta X-ray diffraction scan and the Rietveld refinement of CFO-BFO core-shell nanoparticles. Each peak is assigned to the corresponding Bragg peaks of Fd-3m CFO and R3c BFO phases (denoted with C and B, respectively). (b-g) HAADF-STEM and EDX analyses of the core-shell nanoparticles.

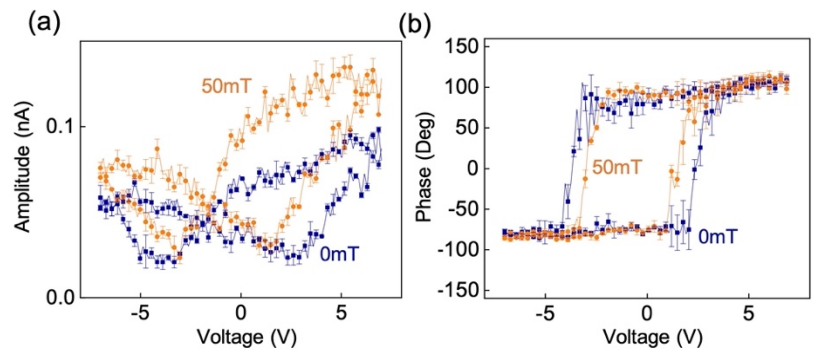

**Figure 2**. Magnetic field dependent local PFM hysteresis (a) amplitude and (b) phase loops.

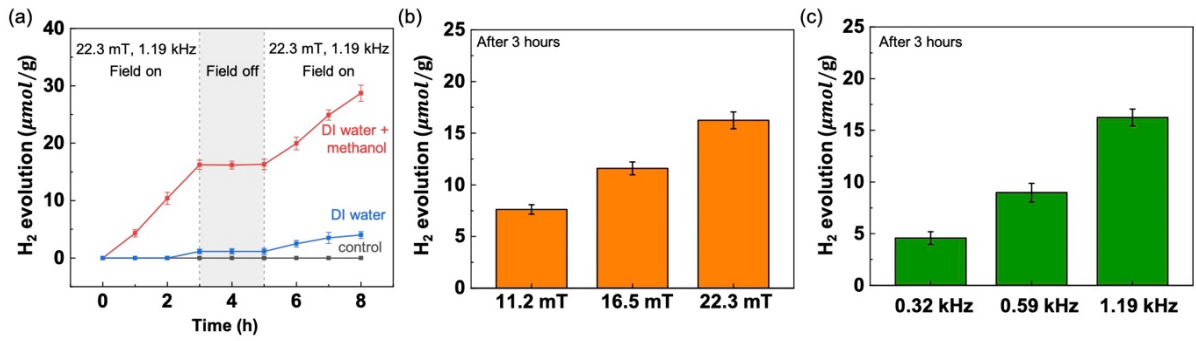

**Figure 3.** (a) Hydrogen evolution measured from gas chromatography (GC) spectra under an alternating magnetic field (22mT, 1.19kHz) as a function of time. Control samples indicate DI water, methanol/DI water solution, CFO nanoparticles mixed with DI water, and CFO nanoparticles mixed with methanol/DI water solution. Magnetic field (b) intensity and (c) frequency dependency of hydrogen evolution (after 3 hours of magnetic field application) in CFO-BFO particles mixed with methanol/DI water solution.

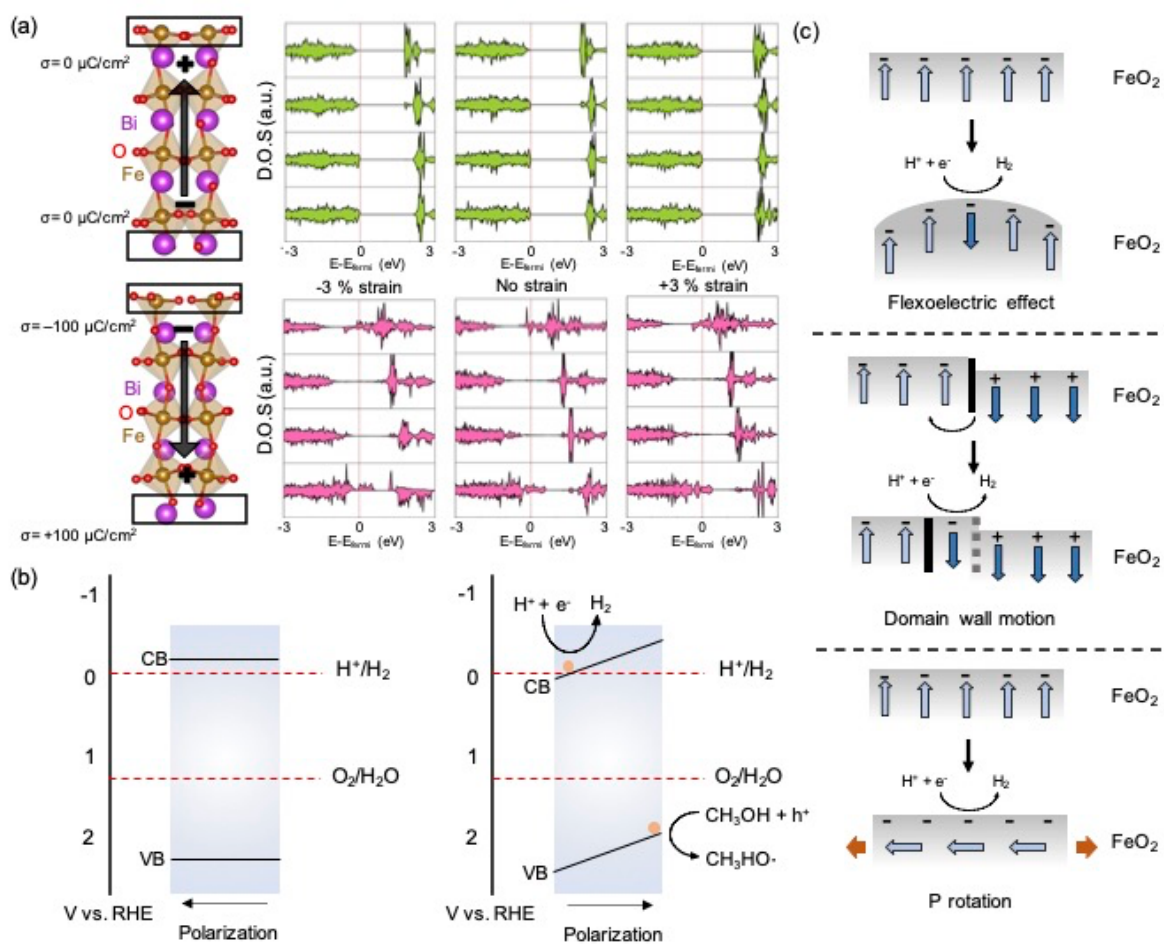

**Figure 4**. (a) BFO (001) slab used for calculation and the calculated layer-by-layer density of states at different strain status for up-polarization (surface-compensated, top) and down-polarization (surface-uncompensated, bottom). (b) Schematic diagram of band bending caused by the reversal of the polarization and resulting electrochemical reactions at the CFO-BFO surfaces. (c) Possible mechanisms for the polarization reversal by magnetoelectric coupling in core-shell nanoparticles. (top) Generation of strain gradients leads to a flexoelectric-induced polarization reversal, (middle) strain-induced motion of domain walls, (bottom) tensile-strain induced in-plane rotation of the polarization.

# Supplementary information

# Magnetoelectric effect in hydrogen harvesting: magnetic field as a trigger of catalytic reactions

Donghoon Kim[a], Ipek Efe[b], Harun Torlakcik[a], Anastasia Terzopoulou[a], Andrea Veciana Picazo[a], Erdem Siringil[a], Fajer Mushtaq[a], Carlos Franco[a], Josep Puigmartí-Luis[d,e], Bradley Nelson[a], Nicola A. Spaldin[b], Chiara Gattinoni*[b,c], Xiang-Zhong Chen*[a], Salvador Pané*[a]

[a] Multi-Scale Robotics Lab, Institute of Robotics and Intelligence Systems, ETH Zürich, Tannenstrasse 3, CH-8092 Zürich, Switzerland. E-mail: chenxian@ethz.ch; vidalp@ethz.ch

[b] Materials Theory Lab, Department of Materials, ETH Zürich, Wolfgang-Pauli-Strasse 27, 8093 Zürich, Switzerland

[c] Department of Chemical and Energy Engineering, London South Bank University, 103 Borough Rd, London SE1 0AA, UK. E-mail: gattinoc@lsbu.ac.uk

[d] Department of Physical Chemistry, University of Barcelona, Martí i Franquès, 1, 08028, Barcelona, Spain

[e] ICREA, Pg. Lluís Companys 23, Barcelona 08010, Spain

## Experimental Methods

Nanoparticles synthesis

$CoFe_2O_4$-$BiFeO_3$ core-shell nanoparticles were synthesized using co-precipitation, hydrothermal and sol-gel process as previously reported [1]. 2g of CTAB was dissolved in 30mL of DI water and afterwards 1g of $FeCl_3 \cdot 6H_2O$ and 0.24 g $CoCl_2$ anhydrous powders were dissolved. Subsequently, 6M NaOH solution was carefully mixed with chemical solution to precipitate CFO nanoparticles. To make CFO nanoparticles single-crystalline, chemical solution was sealed in an autoclave and treated at high temperature for a hydrothermal process. BFO shell was coated by sol-gel process. 0.243g of $Fe(NO_3)_3 \cdot 9H_2O$ and 0.322g of $Bi(NO_3)_3 \cdot 5H_2O$ was dissolved in 60mL of ethylene glycol solution and mixed with 0.1g of as-synthesized CFO nanoparticles. Then the mixture was heated up to 80°C and dried overnight. Dried powder was annealed at 600 °C to crystalize the BFO shell.

Characterization

The crystallinity of synthesized nanoparticles was measured with an X-ray diffractometer (Bruker AXS D8 Advance), equipped with Lynxeye superspeed detector. Transmission electron microscopy (TEM), scanning transmission electron microscopy (STEM) and EDX were performed with FEI Talos F200X. Piezoelectric properties and magnetoelectric coupling were measured with piezoresponse force microscopy (ND-MDT) equipped with an in-plane DC magnetic field setup. An Au-coated conductive tip was used in contact mode in order to apply an alternating voltage and induce piezoelectric oscillations. Local piezoelectric hysteresis loop was measured by applying DC voltage superimposed with small AC voltage. Local piezoelectric hysteresis loop was measured 5 times and averaged. For electrochemical Mott-Schottky analysis, 3mg of CFO-BFO nanoparticles were dispersed in 10 μL Nafion + 10 μL ethanol solution and drop-casted on a conductive ITO glass (cured at 60 °C in a vacuum oven). A carbon rod and Ag/AgCl electrode were used as a counter and reference electrode, respectively, with 0.5M $Na_2SO_4$ electrolyte. For Ag nanoparticles deposition, 10mg of CFO-BFO nanoparticles were dispersed in 10 mL of 50mM $AgNO_3$ solution and 22.3 mT, 1.19kHz AC magnetic field was applied for 1 hour. The particles were then cleaned with Acetone, IPA, and DI water several times and collected for TEM measurements.

Hydrogen evolution measurement

10mg of as-synthesized CFO-BFO core-shell nanoparticles were dispersed in 10mL of DI water and 9mL/1mL of DI water/methanol solution. Magnetoelectrically induced hydrogen evolution reaction was monitored with an online GC setup. The reaction chamber was directly connected to GC (Shimadzu GC 2014) and the pressure of the system was carefully monitored during the reaction. High purity Ar gas was used as both carrier gas and flushing gas. Before applying magnetic fields, the whole system was flushed with Ar for 5 min to remove any other gases remaining in the system. After the reaction, a valve connecting to GC column was opened and 10 µL of gas was injected to GC and analyzed.

DFT calculation

Density functional theory calculations were performed within the periodic supercell approach using the VASP code [2-5]. The optB86b-vdW functional [6], a revised version of the van der Waals (vdW) density functional of Dion et al [7], was chosen for the calculations because it has been shown to describe molecular adsorption on transition metal oxides well [8-10]. Effective on-site interactions for the localized d-orbitals of Fe atoms were considered by adding a Hubbard U term in the Dudarev approach [11] with U-J= 4.0 eV. Core electrons were replaced by projector augmented wave (PAW) potentials [12], while the valence states (5$e^-$ for Bi, 8 $e^-$ for Fe, and 6 $e^-$ for O) were expanded in plane waves with a cut-off energy of 500 eV. The unit cell of the calculated bulk structure has a surface area of $\sqrt{2}a \times \sqrt{2}a$ and height of 2a, where a is the lattice parameter of pseudocubic unit cell. The pseudocubic lattice parameter was calculated to be a=3.95 Å , with the ɣ angle in the rhombohedral structure being ɣ=90.23° by using the optB86b-VdW functional. The difference in the calculated lattice parameters with respect to the experimental structure was below 0.5% [13]. A Monkhorst-Pack k-point grid of (5 x 5 x 1) was used for all calculations. For the density of states calculations, a Monkhorst-Pack k-point grid of 7 x 7 x 1 was used. An antiferromagnetic G-type ordering was imposed, which gave a magnetic moment of 4.15 $\mu_B$ per Fe ion in the bulk BFO. The BFO (001) slabs built for water adsorption had a thickness of four unit cells and were separated from their periodic repetitions in the direction perpendicular to the surface by ~20 Å of vacuum. In our previous work we found that this thickness was sufficient to converge the adsorption energies of the water molecules [14]. A dipole correction along the direction perpendicular to the surface was applied and geometry optimizations were performed with a residual force threshold of 0.01 eV/Å. For investigating the impact of biaxial epitaxial strain on water adsorption, the in-plane lattice parameters of BFO (001) were fixed to corresponding lattice parameter value with the respective applied misfit strain by the relation

$$\text{Misfit } f = \frac{a_{f,strained} - a_{f,relaxed}}{a_{f,strained}}$$

Where $a_{f,strained}$ denotes the strained lattice parameter and $a_{f,relaxed}$ the relaxed bulk lattice parameter. Thus, the negative value of the misfit strain shows compressive strain while the positive value shows the tensile strain. For all different systems a range of from -3% to +3% strain values were used.

Adsorption energies for the water molecules, $E_{ads}$, were calculated as

$$E_{ads} = (E_{water/BFO} - E_{BFO} - n \times E_{water})/n,$$

where $E_{BFO}, E_{water}$, and $E_{water/BFO}$ are the total energies of the relaxed bare slab, an isolated gas phase water molecule, and a system containing n water molecules adsorbed on the slab, respectively. Negative values of the adsorption energy indicate favorable (exothermic) adsorption. Water coverages varying between ½ and 1 monolayer (ML)-where 1 monolayer is one water molecule per surface metal atom- were considered.

|  | $CoFe_2O_4$ | $BiFeO_3$ |
|---|---|---|
| Space group | Fd-3m | R3c |
| a | 8.376 | 5.588 |
| c |  | 13.751 |
| Micro strain | 0 | 0.0058 |
| Rwp | 1.538 ||
| Rexp | 0.956 ||
| $\chi^2$ | 2.58 ||

Table S1. The Rietveld refinement parameters of XRD measurements of CFO-BFO core-shell nanoparticles.

| FeO$_2$ Surface | E$_{ads}$ diss. | E$_{ads}$ intact | BiO Surface | E$_{ads}$ diss. | E$_{ads}$ intact |
|---|---|---|---|---|---|
| -3 % | -0.56 | **-0.76** | -3 % | -0.48 | **-0.60** |
| -1.5 % | -0.49 | **-0.77** | -1.5 % | -0.57 | **-0.64** |
| 0 % | -0.45 | **-0.80** | 0 % | -0.64 | **-0.67** |
| 1.5 % | -0.45 | **-0.85** | 1.5 % | **-0.70** | **-0.70** |
| 3 % | -0.61 | **-0.91** | 3 % | **-0.75** | -0.73 |

Table S2. Adsorption energy of dissociated and intact water molecules on FeO$_2$ and BiO surfaces of BFO slab.

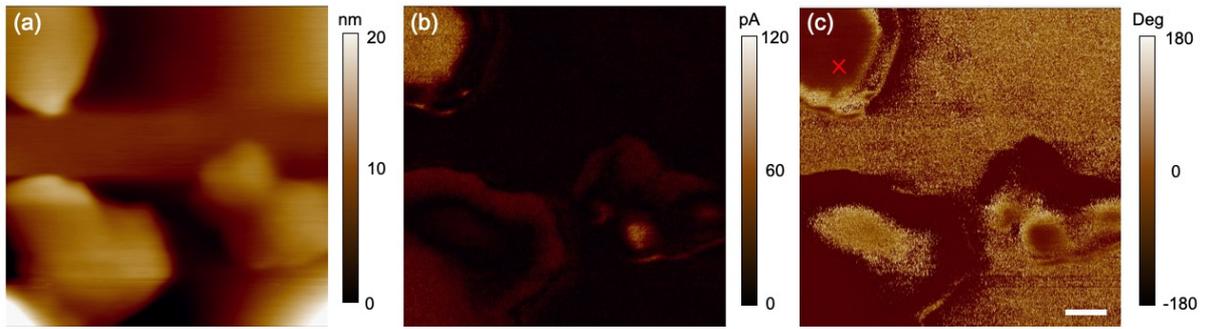

Figure S1. (a) AFM topography, (b) PFM amplitude, and (c) PFM phase images of CFO-BFO core-shell nanoparticles. The point of measurement is indicated with a red cross in (c) and the scale bar indicates 30nm.

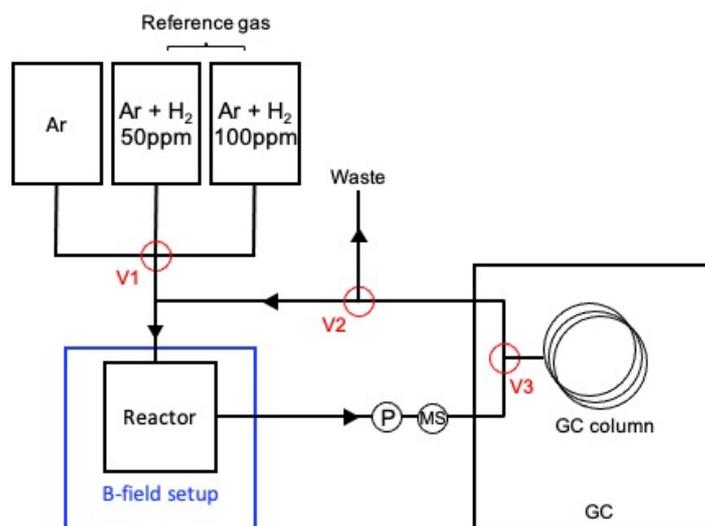

Figure S2. Schematic illustration of the magnetic field assisted online GC setup. The reactor-GC closed system is connected to the gas line and the pressure of the system is maintained constant (with pressure meter (P), 0.15 bar) during the whole reaction time. A Molecular Sieve (MS) was also added to the system to remove water molecules that evaporated from the reactor. By controlling the valve 1 (V1), different reference gases can be flushed into the system. Before the measurement, GC was calibrated with 50ppm and 100 ppm $H_2$ gas mixed with Ar. By opening valve 2 (V2), the setup could be shifted to open system where continuous Ar flushing is possible. Before applying the magnetic field, the system was flushed with Ar gas for 5 min to remove reference gases and other gases in the system. No signal was detected at the beginning of the reaction, indicating only Ar gas existed in the closed online GC system. During measurement, 10 µL of the gas in the system was injected by opening valve 3 (V3) for a certain amount of time and the GC signal was obtained.

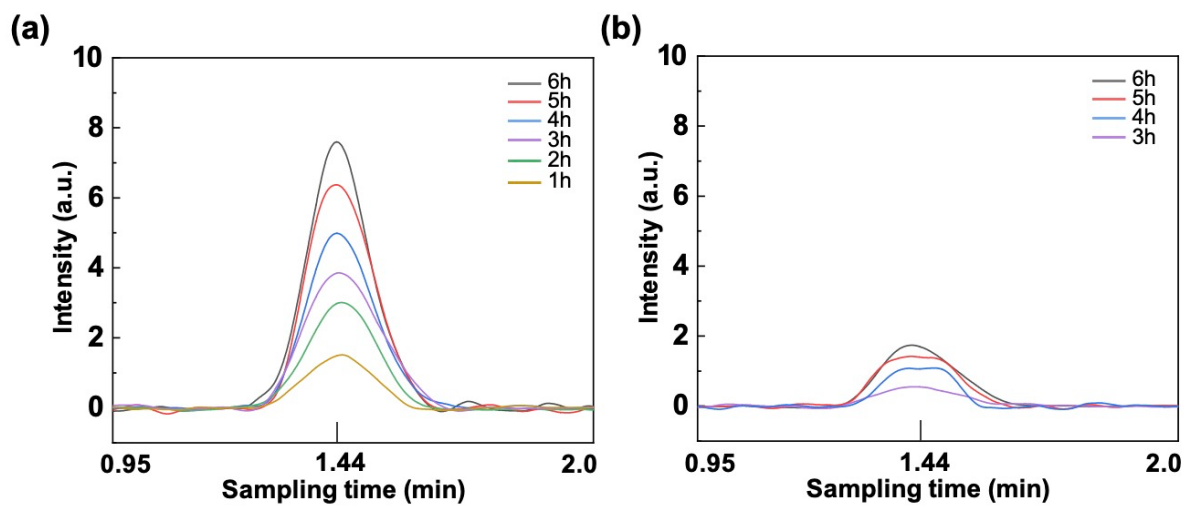

Figure S3. Magnetoelectrically induced hydrogen evolution measured with the gas chromatography of CFO-BFO nanoparticles mixed with (a) DI water/methanol solution and (b) DI water.

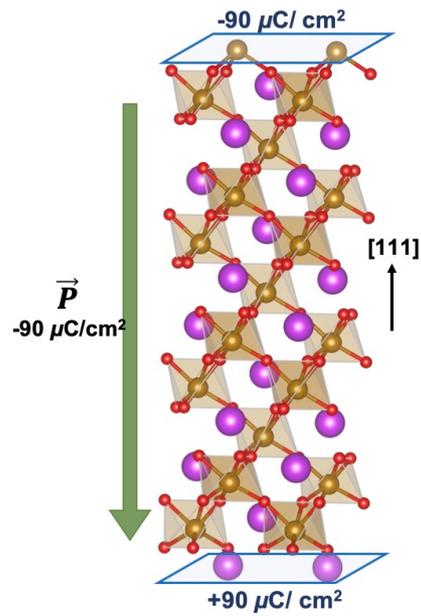

Figure S4. Atomic structures and calculated polarization of the BFO (111) slab. To maintain the charge neutrality of the surface, stable polarization direction points from the positively charged Fe (3+) layer towards the negatively charged $BiO_3$ (-3) surface (downward polarization).

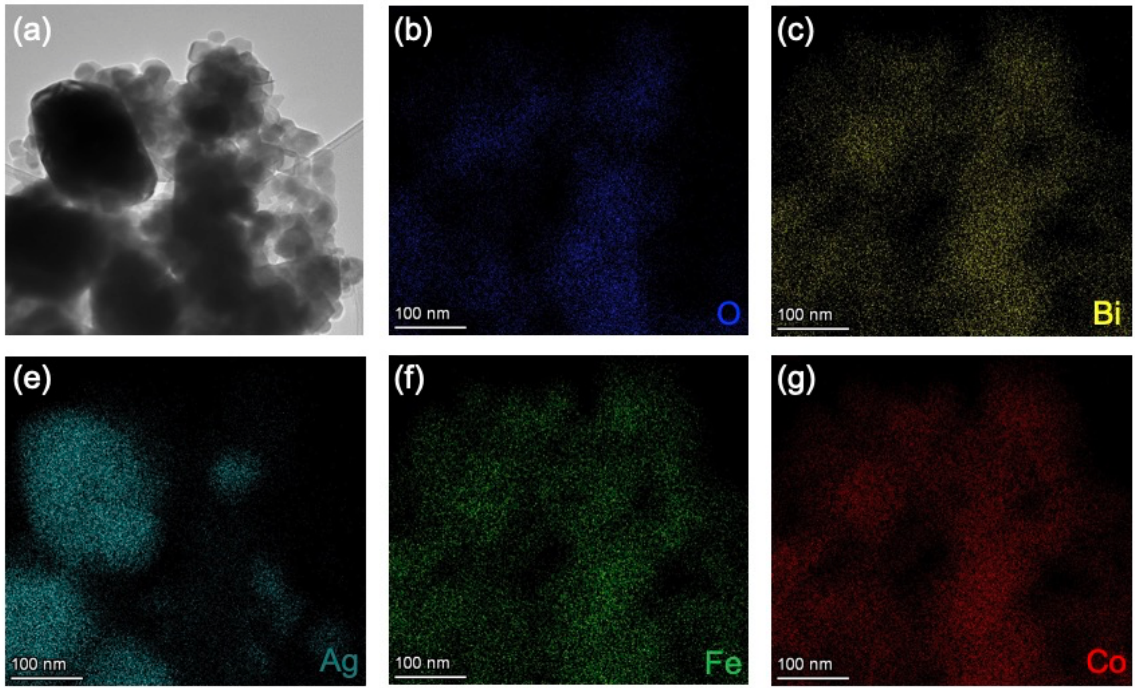

Figure S5. TEM-EDX analysis of magnetoelectrically deposited Ag nanoparticles on CFO-BFO nanoparticles. Ag nanoparticles were deposited around CFO-BFO nanoparticles as can be seen from (a) TEM and (b-g) EDX analyses.

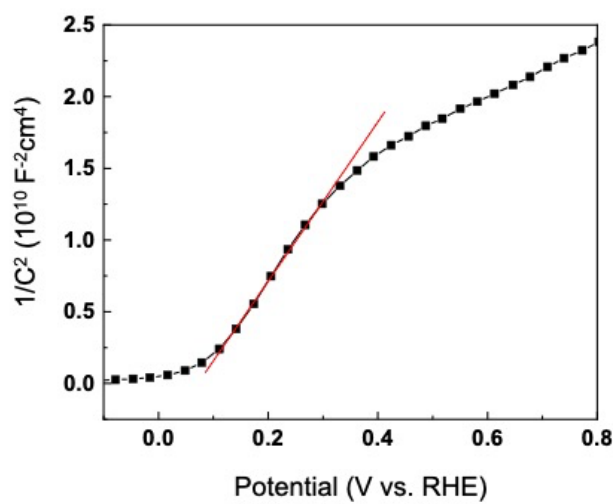

Figure S6. Mott-Schottky plot measured with CFO-BFO nanoparticles. The positive slope indicates an n-type characteristic of CFO-BFO nanoparticles. The electrode potentials were calculated as $E_{RHE} = E_{Ag/AgCl} + 0.059*pH + E°_{Ag/AgCl}$ where $E°_{Ag/AgCl} = 0.205$ V (3.5M KCl), pH = 6.8 (0.5M $Na_2SO_4$ electrolyte). In Mott-Schottky plots, flat band potential ($E_{fb}$) is estimated from the intercept of the linear region ($E_{fb} = 0.095$ V vs. RHE). As the conduction band of the n-type semiconductor is generally about 0.1-0.3 V more negative than the flat band potential [15]; the conduction band of the CFO-BFO nanoparticles are slightly above the redox potential for HER.